\PassOptionsToPackage{usenames,dvipsnames,svgnames,table}{xcolor}
\RequirePackage{fix-cm}
\documentclass[aps,prl,twocolumn,superscriptaddress,groupedaddress]{svjour3}          
\smartqed  

\usepackage[T1]{fontenc}
\usepackage[utf8]{inputenc}
\usepackage[colorlinks=true, linkcolor=Tbluedark, citecolor=Tbluedark, breaklinks=true, bookmarksopen,bookmarksdepth=3]{hyperref}
\usepackage{amsmath}
\usepackage{amssymb}
\usepackage{graphicx}
\usepackage{authblk}
\graphicspath{{images/}}
\usepackage{subcaption}
\usepackage{booktabs}
\usepackage{transparent}
\pdfoutput=1 
\usepackage[english]{varioref}

\usepackage{lineno}

\usepackage{units}
\usepackage[super]{nth}
\usepackage{hepnames}
\newcommand{\PUpsilonFiveS}{\PUpsilon (5 \mathrm{S})}

\newcommand{\BTag}{\PB_{\mathrm{tag}}\xspace}
\newcommand{\BSig}{\PB_{\mathrm{sig}}\xspace}
\newcommand{\FEI}{\texttt{FEI}\xspace}
\newcommand{\FR}{\texttt{FR}\xspace}

\usepackage[numbers]{natbib}

\usepackage{xcolor}
\usepackage{tikz}
\usetikzlibrary{calc}
\usetikzlibrary{chains}
\usetikzlibrary{positioning,arrows}
\usetikzlibrary{decorations.pathmorphing}
\usetikzlibrary{decorations.markings}
\usetikzlibrary{fadings,positioning,through}

\newcommand*{\belowrulesepcolor}{%
  \noalign{%
    \kern-\belowrulesep
    \begingroup
      \color{HeadRowColor}%
      \hrule height\belowrulesep
    \endgroup
  }%
}
\newcommand*{\aboverulesepcolor}{%
  \noalign{%
    \begingroup
      \color{HeadRowColor}%
      \hrule height\aboverulesep
    \endgroup
    \kern-\aboverulesep
  }%
}

\usepackage{listings}
\definecolor{Tblue}{HTML}{3465A4}	
\definecolor{Tbluedark}{HTML}{204A87}	
\definecolor{Tbluelight}{HTML}{729FCF}	
\definecolor{Tbluelighter}{HTML}{8CC4FF}	
\definecolor{Tbrown}{HTML}{C17D11}	
\definecolor{Tbrowndark}{HTML}{8F5902}	
\definecolor{Tbrownlight}{HTML}{E9B96E}	
\definecolor{Tgray}{HTML}{888A85}	
\definecolor{Tgraydark}{HTML}{555753}	
\definecolor{Tgraydarker}{HTML}{2E3436}	
\definecolor{Tgraylight}{HTML}{BABDB6}	
\definecolor{Tgraylight2}{HTML}{E4E6E2}	
\definecolor{Tgraylight3}{HTML}{F0F2EE}	
\definecolor{Tgreen}{HTML}{73D216}	
\definecolor{Tgreendark}{HTML}{4E9A06}	
\definecolor{Tgreenlight}{HTML}{8AE234}	
\definecolor{Tred}{HTML}{CC0000}	
\definecolor{Treddark}{HTML}{A40000}	
\definecolor{Tredlight}{HTML}{EF2929}	
\definecolor{Tlilac}{HTML}{75507B}	
\definecolor{Tlilacdark}{HTML}{5C3566}	
\definecolor{Tlilaclight}{HTML}{AD7FA8}	
\definecolor{Tyellow}{HTML}{EDD400}	
\definecolor{Tyellowdark}{HTML}{C4A000}	
\definecolor{Tyellowlight}{HTML}{FCE94F}
\definecolor{Torange}{HTML}{F57900}	
\definecolor{Torangedark}{HTML}{CE5C00}	
\definecolor{Torangelight}{HTML}{FCAF3E}

\definecolor{Tgrayforkitprinter}{HTML}{EEEEEE}	
\colorlet{HeadRowColor}{Tgrayforkitprinter}
\colorlet{CrossFeedColor}{Tblue}

\begin{document}

\title{The Full Event Interpretation}
\subtitle{An exclusive tagging algorithm for the Belle~II experiment} 

\author[1]{T.~Keck}
\author[2]{F.~Abudin\'en}
\author[1]{F.U.~Bernlochner}
\author[3]{R.~Cheaib}
\author[4]{S.~Cunliffe}
\author[1]{M.~Feindt}
\author[4]{T.~Ferber}
\author[1]{M.~Gelb}
\author[1]{J.~Gemmler}
\author[1]{P.~Goldenzweig}
\author[1]{M.~Heck}
\author[5]{S.~Hollitt}
\author[6]{J.~Kahn}
\author[7]{J-F.~Krohn}
\author[6]{T.~Kuhr}
\author[4]{I.~Komarov}
\author[2]{L.~Ligioi}
\author[8]{M.~Lubej}
\author[1]{F.~Metzner}
\author[1]{M.~Prim}
\author[1]{C.~Pulvermacher}
\author[6]{M.~Ritter}
\author[1]{J.~Schwab}
\author[1]{W.~Sutcliffe}
\author[9]{U.~Tamponi}
\author[4]{F.~Tenchini}
\author[10]{N.~E.~Toutounji}
\author[7]{P.~Urquijo}
\author[1]{D.~Weyland}
\author[8]{A.~Zupanc}

\affil[1]{Karlsruhe Institute of Technology, Karlsruhe, Germany}
\affil[2]{Max-Planck-Institut f\"ur Physik, Munich, Germany}
\affil[3]{University of Mississippi, Mississippi, USA}
\affil[4]{Deutsches Elektronen-Synchrotron, Hamburg, Germany}
\affil[5]{University of Adelaide, Adelaide, Australia}
\affil[6]{Ludwig Maximilians Universit\"at, Munich, Germany}
\affil[7]{University of Melbourne, Melbourne, Australia}
\affil[8]{Jo\v{z}ef Stefan Institute, Ljubljana, Slovenia}
\affil[9]{INFN - Sezione di Torino, Torino, Italy}
\affil[10]{University of Sydney, Sydney, Australia}

\institute{T. Keck \at
			Karlsruher Institut für Technologie, Campus Süd\\
			Institut für Experimentelle Teilchenphysik\\
		    Wolfgang-Gaede-Str. 1 \\
		    76131 Karlsruhe
            \email{thomas.keck2@kit.edu}
}

\titlerunning{FEI}
\authorrunning{FEI}

\date{Received: date / Accepted: date}

\maketitle 

\begin{abstract}
The Full Event Interpretation is presented: a new exclusive tagging algorithm used by the high-energy physics experiment Belle~II.
The experimental setup of Belle~II allows the precise measurement of otherwise inaccessible $\PB$ meson decay-modes.
The Full Event Interpretation algorithm enables many of these measurements.
The algorithm relies on machine learning to automatically identify plausible $\PB$ meson decay chains based on the data recorded by the detector.
Compared to similar algorithms employed by previous experiments, the Full Event Interpretation provides a greater efficiency,
yielding a larger effective sample size usable in the measurement.
\end{abstract}

\section{Introduction}

The Belle~II experiment is located at the SuperKEKB electron-positron collider in Tsukuba, Japan, and was commissioned in 2018.
The experiment is designed to perform a wide range of high-precision measurements in all fields of heavy flavour physics, in particular
it will investigate the decay of $\PB$ mesons \citep{B2TDR}. For this purpose, the experiment is expected to record about $40$ billion collision events each containing an $\PUpsilonFourS$ resonance, which at least $96 \%$ of the time decays into exactly \textbf{two} $\PB$ mesons (a $\PB \APB$ pair).
Each $\PB$ meson decays via various intermediate states into a set of final-state particles, which are considered
stable in the Belle~II detector. In general, charged final-state particles are reconstructed as tracks in the central drift
chamber and in the inner silicon-based vertex detectors, whereas neutral final-state particles are reconstructed as energy depositions (called clusters) in the electromagnetic calorimeter.
The entire experimental setup of the detector and the collider is described in more detail in \citet{B2TDR}.

\clearpage

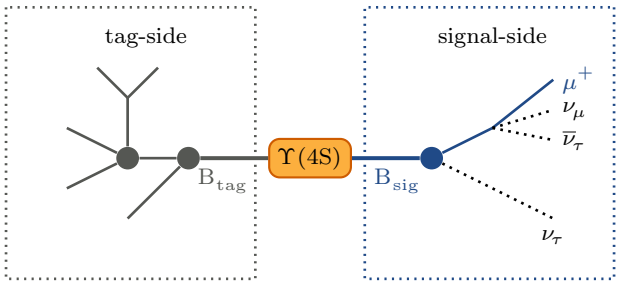
\begin{figure}
	\centering
\begin{tikzpicture}[%
scale=0.8,
bstage/.style={
	rectangle,
	draw=Torangedark, 
	thick,
	font=\small,
	fill=Torangelight,
	align=center,
	rounded corners,
	minimum height=1em
},   
decaypoint/.style={
	circle,
	align=center,
},   
]
	\draw (0,0) node[bstage] (Y) {$\PUpsilonFourS$};
	\node[decaypoint,fill=Tgraydark] at (-2,0) (BM) {};
	\node[decaypoint,fill=Tbluedark] at (2,0) (BP) {};
	
	\coordinate (BMD0K) at (-4,-0.5);
	\coordinate (BMD0P) at (-4,0.5);
	\coordinate (BMD0PDP) at (-3.5,0.25);
	\coordinate (BMD0PTP) at (-3.5,0.75);
	\coordinate (BMD0KS) at (-3,1.0);
	\coordinate (BMD0KSp1) at (-3.5,1.5);
	\coordinate (BMD0KSp2) at (-2.5,1.5);
	\coordinate (BMPM) at (-3,-1);
	
	\coordinate (BPP) at (4,1);
	\coordinate (BPN) at (4,-1);
	\coordinate (BPL) at (3,0.5);

	\coordinate (BPMU) at (4.0,1.3);
	\coordinate (BPNM) at (4,0.8);
	\coordinate (BPNT) at (4.0,0.3);
	
	\draw[line width=1.5pt,Tgraydark] (Y.west) -> node[below,xshift=-1.5mm] { $\BTag$ } (BM) ;
	\draw[line width=1.5pt,Tbluedark] (Y.east) -> node[below,xshift=1.5mm] { $\BSig$ } (BP);
	
	\draw[line width=1pt,black,dotted] (BP) -> (BPN) node[below] { $\Pnut$ };
	\draw[line width=1pt,Tbluedark] (BP) -> (BPL) node[right] {};
	\draw[line width=1pt,Tbluedark] (BPL) -> (BPMU) node[right] { $\APmuon$ };
	\draw[line width=1pt,black,dotted] (BPL) -> (BPNM) node[right] { $\Pnum$ };
	\draw[line width=1pt,black,dotted] (BPL) -> (BPNT) node[right] { $\APnut$ };
	
	\draw[Tbluedark,thick,dotted,line width=1pt] ($(BPP)+(1.0,1.5)$)  rectangle ($(BPN)+(-3.1,-1.0)$);
	\draw ($(BPP)+(-1,+1.0)$) node[align=right] {signal-side};
	
	\node[decaypoint,fill=Tgraydark] at (-3,0) (BMD0) {};
	\draw[line width=1pt,Tgraydark] (BM) -> (BMD0);
	\draw[line width=1pt,Tgraydark] (BMD0) -> (BMD0K);
	\draw[line width=1pt,Tgraydark] (BMD0) -> (BMD0P);
	\draw[line width=1pt,Tgraydark] (BM) -> (BMPM);
	\draw[line width=1pt,Tgraydark] (BMD0) -> (BMD0KS);
	\draw[line width=1pt,Tgraydark] (BMD0KS) -> (BMD0KSp1);
	\draw[line width=1pt,Tgraydark] (BMD0KS) -> (BMD0KSp2);
	
	\draw[Tgraydark,thick,dotted,line width=1pt] ($(BMD0P)+(-1.0,2.0)$)  rectangle ($(BMPM)+(2.1,-1.0)$);
	\draw ($(BMD0P)+(+1.3,1.5)$) node[align=left] {tag-side};

\end{tikzpicture}
\caption{Schematic overview of a $\PUpsilonFourS$ decay: (Left) a common tag-side decay $\BTag^- \rightarrow \PDzero (\rightarrow \PKshort (\rightarrow \Ppiminus \Ppiplus)  \Ppiminus \Ppiplus) \Ppiminus$ and
(right) a typical signal-side-decay $\BSig^+ \rightarrow \APtauon (\rightarrow \APmuon \Pnum \APnut) \Pnut$. The two sides overlap spatially in the detector, therefore the assignment of a measured track to one of the sides is not known a priori.}
\label{fig:fei:example_decay}
\end{figure}

The measurement of the branching fraction of rare decays like $\PB \rightarrow \Ptau \Pnu$, $\PB \rightarrow \PK \Pnu \Pnu$ or $\PB \rightarrow \Pl \Pnu \Pgamma$, with undetectable neutrinos in their final states, is challenging.
However, the second $\PB$ meson in each event can be used to constrain the allowed decay chains.
This general idea is known as \textbf{tagging}.
Conceptually, each $\PUpsilonFourS$ event is divided into two sides: The signal-side containing
the tracks and clusters compatible with the assumed signal $\BSig$ decay the physicist is interested in,
e.g. a rare decay like $\PB \rightarrow \Ptau \Pnu$;
and the tag-side containing the remaining tracks and clusters compatible with an arbitrary $\BTag$ meson decay.
\autoref{fig:fei:example_decay} depicts this situation.

The initial four-momentum of the produced $\PUpsilonFourS$ resonance is precisely known and no additional particles
are produced in this primary interaction.
Therefore, because of the relevant quantum numbers conservation, knowledge about the properties of the tag-side $\BTag$ meson allows one to recover information about
the signal-side $\BSig$ meson which would otherwise be inaccessible.
Most importantly, all reconstructed tracks and clusters which are not assigned to the $\BTag$ mesons must be compatible
with the signal-decay of interest.

Ideally, a \textbf{full} reconstruction of the entire \textbf{event} has to take all reconstructed tracks
and clusters into account to attain a correct \textbf{interpretation} of the measured data.
The \texttt{\textbf{Full Event Interpretation}} (\FEI) algorithm presented in this article is a new exclusive tagging algorithm
developed for the Belle~II experiment, embedded in the Belle II Analysis Software Framework (basf2) \citep{BASF2}. The \FEI automatically constructs plausible 
$\BTag$ meson decay chains compatible with the observed tracks and clusters,
and calculates for each decay chain the probability of it correctly describing the true process using gradient-boosted decision trees. ``Exclusive'' refers to the reconstruction of a particle (here the $\BTag$) assuming an explicit decay channel.

Consequently, exclusive tagging reconstructs the $\BTag$ independently of the
$\BSig$ using either \textbf{hadronic} or \textbf{semileptonic} $\PB$ meson decay channels.
The decay chain of the $\BTag$ is explicitly reconstructed
and therefore the assignment of tracks and clusters to the tag-side and signal-side
is known.

In the case of a measurement of an exclusive branching fraction like $\BSig \rightarrow \Ptau \Pnut$, the 
entire decay chain of the $\PUpsilonFourS$ is known.
As a consequence, all tracks and clusters measured by the detector should
be already accounted for.
In particular, the requirement of no additional tracks, besides the
ones used for the reconstruction of the $\PUpsilonFourS$, is an extremely 
powerful and efficient way to remove most reducible\footnote{Reducible background
has distinct final-state products from the signal.} backgrounds.
This requirement is called the \textbf{completeness constraint} throughout this text.

In the case of a measurement of an inclusive branching fraction like $\BSig \rightarrow X_{\Pqu} \Pl \nu$,
all remaining tracks and clusters, besides the ones used for the lepton $\Pl$ and the $\BTag$ meson, are identified with the $X_{\Pqu}$ system.
Hence, the branching fraction can be determined without explicitly assuming a decay chain for the $X_{\Pqu}$ system.

The performance of an exclusive tagging algorithm depends on the \textbf{tagging efficiency} (i.e. the fraction of $\PUpsilonFourS$ events which can be tagged),
the \textbf{tag-side efficiency} (i.e. the fraction of $\PUpsilonFourS$ events with a correct tag) and
on the quality of the recovered information, which determines the \textbf{tag-side purity} (i.e. the fraction of the tagged $\PUpsilonFourS$ events with a correct tag) of the tagged events.

The exclusive tag typically provides a pure sample (i.e. purities up to $90 \%$ are possible). 
But this approach suffers from a low tag-side efficiency, just a few percent, since
only a tiny fraction of the $\PB$ decays can be explicitly reconstructed due
to the large amount of possible decay channels and their high multiplicity. 
The imperfect reconstruction efficiency of tracks and clusters further degrades the
efficiency. 

Both the quality of the recovered information and the systematic uncertainties
depend on the decay channel of the $\BTag$,
therefore we distinguish further between hadronic and semileptonic
exclusive tagging.

Hadronic tagging considers only hadronic $\PB$ decay chains for the tag-side \citep[Section 7.4.1]{PhysicsOfTheBFactories}. Hence,
the four-momentum of the $\BTag$ is well-known and the tagged sample
is very pure.
A typical hadronic $\PB$ decay has a branching fraction of $\mathcal{O}(10^{-3})$.
As a consequence, hadronic tagging suffers from a low tag-side efficiency and
can only be applied to a tiny fraction of the recorded events. Large combinatorics 
of high-multiplicity decay channels further complicate the reconstruction and require
tight selection criteria. 

Semileptonic tagging considers only semileptonic $\PB \rightarrow \PD \Pl \Pnu$ and $\PB \rightarrow \PDst \Pl \Pnu$  decay channels \citep[Section 7.4.2]{PhysicsOfTheBFactories}.
Due to the presence of a high-momentum lepton these decay channels can be easily identified and the semileptonic tagging
usually yields a higher tag-side efficiency compared to hadronic tagging due to the large semileptonic branching fractions.
On the other hand, the semileptonic tag will miss kinematic
information due to the neutrino in the final state of the decay. Hence, the sample is
not as pure as in the hadronic case.

To conclude, the \FEI provides a hadronic and semileptonic tag for $\PBpm$ and $\PBz$ mesons.
This enables the measurement of exclusive decays with several neutrinos and inclusive decays.
In both cases the \FEI provides an explicit tag-side decay chain with an associated probability.

\section{Method}
\label{sec:method}
\begin{figure}
\centering
\begin{tikzpicture}[%
scale=0.75,
    detector/.style={
      rectangle,
      draw=Tgray,
      thick,
      font=\small,
      fill=Tgraylight2,
      postaction={path fading=north, fading angle=-45, fill=Tgraylight},
      text width=10em,
      align=center,
      rounded corners,
      minimum height=2em
    },  
    fsp/.style={
      rectangle,
      draw=Tblue!25!Tgray,
      thick,
      font=\small,
      fill=Tbluelighter!25!Tgraylight2,
      postaction={path fading=north, fading angle=-45, fill=Tbluelight!25!Tgraylight},
      text width=2em,
      align=center,
      rounded corners,
      minimum height=2em
    },  
    greenfsp/.style={
      rectangle,
      draw=Tblue!50!Tgray,
      thick,
      font=\small,
      fill=Tbluelighter!50!Tgraylight2,
      postaction={path fading=north, fading angle=-45, fill=Tbluelight!50!Tgraylight},
      text width=2em,
      align=center,
      rounded corners,
      minimum height=2em
    },  
    dstage/.style={
      rectangle,
      draw=Tblue!75!Tgray,
      thick,
      font=\small,
      fill=Tbluelighter!75!Tgraylight2,
      postaction={path fading=north, fading angle=-45, fill=Tbluelight!75!Tgraylight},
      text width=6em,
      align=center,
      rounded corners,
      minimum height=2em
    },  
    bstage/.style={
      rectangle,
      draw=Tbluedark,
      thick,
      font=\small,
      fill=Tbluelighter,
      postaction={path fading=north, fading angle=-45, fill=Tbluelight},
      text width=4em,
      align=center,
      rounded corners,
      minimum height=1em
    },  
  ]

\draw (1.0,1) node[detector, text width=4em] (Tr) {\texttt{Tracks}};
\draw (4.3,1) node[detector,text width=9em] (Vo) {\texttt{Displaced Vertices}};
\draw (8.4,1) node[detector,text width=8em] (Cl) {\texttt{Neutral Clusters}};

\draw (9,-2) node[greenfsp] (Pio) {$\Ppizero$};

\draw (6.8,-0.5) node[fsp] (Kaol) {$\PKlong$};
\draw[Tgraydark] (Cl.south) -> (Kaol.north);

\draw (3.8,-3) node[greenfsp] (Kao) {$\PKshort$};
\draw[Tgraydark] (Vo.south) -> (Kao.north);
\draw (4.6,-0.5) node[fsp] (Pi) {$\Ppiplus$};
\draw[Tbluedark!25!Tgraydark] (Pi.south) -> (Kao.north);
\draw[Tbluedark!50!Tgraydark] (Pio.west) -> (Kao.north);

\draw (0.4,-0.5) node[fsp] (El) {$\Pep$};
\draw (1.8,-0.5) node[fsp] (Mu) {$\APmuon$};
\draw (3.2,-0.5) node[fsp] (Ka) {$\PKplus$};
\draw[Tgraydark] (Tr.south) -> (El.north);
\draw[Tgraydark] (Tr.south) -> (Mu.north);
\draw[Tgraydark] (Tr.south) -> (Pi.north);
\draw[Tgraydark] (Tr.south) -> (Ka.north);

\draw (9.8,-0.5) node[fsp] (Ga) {$\Pgamma$};

\draw[Tgraydark] (Vo.south) -> (Ga.north);
\draw[Tgraydark] (Cl.south) -> (Ga.north);
\draw[Tbluedark!25!Tgraydark] (Ga.west) -> (Pio.north);

\draw (9.0,-5) node[dstage] (DStar) {$\PDst^0 \ \PDst^+ \ \PDsstar$};

\draw[Tbluedark!25!Tgraydark] (Pi.south) -> (DStar.north);
\draw[Tbluedark!50!Tgraydark] (Pio.south) -> (DStar.north);
\draw[Tbluedark!25!Tgraydark] (Ga.south) -> (DStar.north);

\draw (6.25,-6) node[bstage] (B) {$\PBzero\ \PBplus$};
\draw[Tbluedark!25!Tgraydark] (Ka.south) -> (B.north);
\draw[Tbluedark!25!Tgraydark] (Pi.south) -> (B.north);
\draw[Tbluedark!25!Tgraydark] (Kaol.south) -> (B.north);
\draw[Tbluedark!50!Tgraydark] (Pio.south) -> (B.north);
\draw[Tbluedark!50!Tgraydark] (Kao.south) -> (B.north);

\draw[Tbluedark!25!Tgraydark] (El.south) -> (B.north);
\draw[Tbluedark!25!Tgraydark] (Mu.south) -> (B.north);

\draw[Tbluedark!50!Tgraydark] (DStar.south) -> (B.north);

\draw (6.25,-4) node[dstage] (D) {$\PDzero\ \PDplus\ \PDs$};
\draw[Tbluedark!25!Tgraydark] (Ka.south) -> (D.north);
\draw[Tbluedark!25!Tgraydark] (Pi.south) -> (D.north);
\draw[Tbluedark!25!Tgraydark] (Kaol.south) -> (D.north);
\draw[Tbluedark!25!Tgraydark] (El.south) -> (D.north);
\draw[Tbluedark!25!Tgraydark] (Mu.south) -> (D.north);
\draw[Tbluedark!50!Tgraydark] (Kao.south) -> (D.north);
\draw[Tbluedark!50!Tgraydark] (Pio.south) -> (D.north);
\draw[Tbluedark!50!Tgraydark] (D.east) -> (DStar.north);
\draw[Tbluedark!50!Tgraydark] (D.south) -> (B.north);

\draw (1.75,-2) node[greenfsp] (J) {$\PJpsi$};
\draw[Tbluedark!25!Tgraydark] (El.south) -> (J.north);
\draw[Tbluedark!25!Tgraydark] (Mu.south) -> (J.north);
\draw[Tbluedark!50!Tgraydark] (J.south) -> (B.north);

\draw (3.8,-3) node[greenfsp] (Kao) {$\PKshort$};
\end{tikzpicture}
\caption{Schematic overview of the \FEI. The algorithm operates on objects identified by the reconstruction software of the Belle~II detectors:
charged tracks, neutral clusters and displaced vertices.
In six distinct stages, these basics objects are interpreted as final-state particles ($\Pep$, $\APmuon$, $\PKplus$, $\Ppiplus$, $\PKlong$, $\Pgamma$)
combined to form intermediate particles ($\PJpsi$, $\Ppizero$, $\PKshort$, $\PD$, $\PDstar$) and finally form the tag-side $\PB$ mesons.}
\label{fig:fei:fei}
\end{figure}
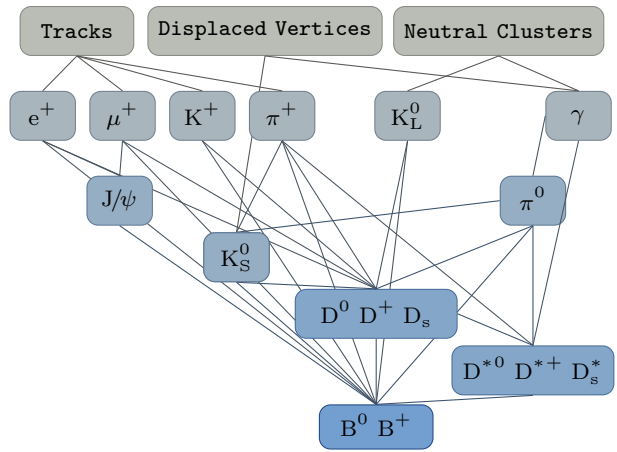

The \FEI algorithm follows a hierarchical approach with six stages, visualized in \autoref{fig:fei:fei}.
Final-state particle candidates are constructed using the reconstructed tracks and clusters,
and combined to intermediate particles until the final $\PB$ candidates are formed. 
The probability of each candidate to be correct is estimated by a multivariate classifier.
A multivariate classifier maps a set of input features (e.g. the four-momentum or the vertex position) to a real-valued output, which can be
interpreted as a probability estimate.
The multivariate classifiers are constructed by optimizing a loss-function (e.g. the mis-classification rate) on Monte Carlo simulated $\PUpsilonFourS$ events and are described later in detail. 

All steps in the algorithm are configurable. Therefore, the decay channels used, the cuts employed, the choice of the input features, and hyper-parameters of the multivariate classifiers depend on the configuration.
A more detailed description of the algorithm and the default configuration can be found in \citet{feithesis} and
in the following we give a brief overview over the key aspects of the algorithm.

\subsection{Combination of Candidates}
\label{subsec:fei:reconstruction}
Charged final-state particle candidates are created from tracks assuming different particle hypotheses.
Neutral final-state particle candidates are created from clusters and displaced vertices constructed by oppositely charged tracks.
Each candidate can be correct (signal) or wrong (background).
For instance, a track used to create a $\Ppiplus$ candidate can originate from a pion traversing
the detector (signal), from a kaon traversing the detector (background) or originates from a random combination of hits
from beam-background (also background). 

All candidates available at this stage are combined to intermediate particle candidates in the subsequent stages,
until candidates for the desired $\PB$ mesons are created.
Each intermediate particle has multiple possible decay channels, which can be used to create
valid candidates.
For instance, a $\PBm$ candidate can be created by combining a $\PDzero$ and a $\Ppiminus$ candidate,
or by combining a $\PDzero$, a $\Ppiminus$ and a $\Ppizero$ candidate.
The $\PDzero$ candidate could be created from a $\PKminus$ and a $\Ppiplus$,
or from a $\PKshort$ and a $\Ppizero$.

The \FEI reconstructs more than $100$ explicit decay channels,
leading to $\mathcal{O}(10000)$ distinct decay chains.

\subsection{Multivariate Classification}
\label{subsec:fei:multivariate_classification}
The \FEI employs multivariate classifiers to estimate the probability of each candidate to be correct, which can be used to discriminate correctly identified candidates from background. For each final-state particle and for each decay channel of an intermediate particle, a multivariate classifier is trained which estimates the signal probability that the candidate is correct. In order to use all available information at each stage, a network of multivariate classifiers is built, following the hierarchical structure in \autoref{fig:fei:fei}.

For instance, the classifier for the decay of $\PBm \rightarrow \PDzero \Ppiminus$ would
use the signal probability of the $\PDzero$ and $\Ppiminus$ candidates, as input features to estimate
the signal probability of the $\PBm$ candidate created by combining the aforementioned $\PDzero$ and $\Ppiminus$ candidates.

Additional input features of the classifiers are the kinematic and vertex fit information of the candidate and its daughters.
The multivariate classifiers used by the \FEI are trained on Monte Carlo simulated events.
The training is fully automatized and distributed using a map-reduce approach \citep{mapReduce}. 
Monte Carlo simulated data used to train the FEI is partitioned. 
At each reconstruction stage the partitioned data is distributed to nodes where the reconstruction is performed and training datasets are produced (the mapping stage). 
The reduction stage consists of merging the training datasets and training multivariate classifiers with these training datasets.

The available information flows from the data provided by the detector through the intermediate candidates into the final $\PB$ meson candidates, yielding a single number which can be used to distinguish correctly 
from incorrectly identified $\BTag$ mesons. The process is visualized in \autoref{fig:fei:fei}. 
This allows one to tune the trade-off between tag-side efficiency and tag-side purity of the algorithm by requiring a minimal signal probability.
By contrast, most exclusive measurements by Belle, which used the previous \FR algorithm, 
chose a working point near the maximum tag-side efficiency as described in \autoref{sec:previous_work}.

\subsection{Combinatorics}
\label{subsec:fei:combinatorics}
It is not feasible to consider all possible $\PB$ meson candidates created by all possible combinations.
The amount of possible combinations scales with the factorial in the number of tracks and clusters.
This problem is known as \textbf{combinatorics} in high-energy physics.
Furthermore, it is not worthwhile to consider all possible $\PB$ meson candidates, because
all of them are wrong except for two in the best-case scenario.

The \FEI uses two sets of so-called \textbf{cuts}. A cut is a criterion that a candidate has to fulfill
to be considered further. For instance, one could demand that the beam-constrained mass of the $\PB$ meson candidate
is near the nominal mass $\unit[5.28]{GeV}$ of a $\PB$ meson particle, or that a $\APmuon$ candidate
has a high muon particle identification likelihood, which combines sub-detector information to identify muons.

Directly after the creation of the candidate (either from a track/cluster, or by combining other candidates),
but before the application of the multivariate classifier, the \FEI uses loose and fast \textbf{pre-cuts}
to remove wrongly identified candidates (background), without losing signal.
The main purpose of these cuts is to save computing time and to reduce the memory consumption.
These \textbf{pre-cuts} are applied separately for each decay channel.

At first, a very loose fixed cut is applied on a quantity which is fast to calculate e.g. the energy for photons, the invariant mass for $\PD$ mesons, the energy released in the decay for $\PDst$ mesons, or the beam-constrained mass for hadronic $\PB$ mesons.
Secondly, the remaining candidates are ranked according to a quantity, which is fast to calculate (usually the same quantity as above is used here).
Only the $n$ (usually between $10$ and $20$) best-candidates in each decay channel are further considered, the others are discarded.
This best-candidate selection ensures that each decay channel and each event receives roughly the same amount of computing time.

Next, the computationally expensive parts of the reconstruction are performed on each candidate: the matching of the reconstructed candidates to the generated particles (in case of simulated events), the vertex fitting, and the multivariate classification.

After the multivariate classifiers have estimated the signal probability of each candidate, the candidates of
different decay channels can be compared.
Here the \FEI uses tighter \textbf{post-cuts} to aggressively remove incorrectly reconstructed candidates using all available information.
The main purpose of these cuts is to restrict the number of candidates per particle to a manageable number.

At first, there is a loose fixed cut on the signal probability, to remove unreasonable candidates.
Secondly, the remaining candidates are ranked according to their signal probability.
Only the $m$ (usually between $10$ and $20$) best-candidates of the particle (i.e. over all decay channels) are further considered,
the others are discarded.
This best-candidate selection ensures that the amount of candidates produced in the next stage is reasonably low and can be handled by the computing system.

\subsection{Performance}
\label{subsec:fei:performance}
Applying the \FEI to $\mathcal{O}(1 \, \text{billion})$ events is a CPU-intensive task.
An optimized runtime and a small memory-footprint are key for a practical application and save computing resources.
The \FEI spends most CPU time on vertex fitting ($38 \%$), particle combination ($27 \%$), and classifier inference ($15 \%$).
All three tasks have been carefully optimized.

The \FEI uses only a fast and simple unconstrained vertex fit during the reconstruction, and feeds the calculated information into its multivariate classifiers.
The user can refit the whole decay chain of the final B candidates, including mass and/or interaction point profile constraints if desired.
A dedicated fitter (called \texttt{FastFit}) based on a Kalman Filter \citep{Fruhwirth} was implemented for the \FEI, which requires drastically less computing time than the default implementation used by Belle~II and yields very similar results. Due to this fitter an overall speedup of the \FEI of $2.74$ was observed.
The \texttt{FastFit} code is licensed under GPLv3 and available on GitHub \citep{FastFitRepo}.

As explained in \autoref{subsec:fei:combinatorics}, the number of candidates
which have to be processed scales as the factorial of the multiplicity of the channel.
In previous approaches the runtime and the maximum memory consumption was dominated by a few high-multiplicity events
and tight cuts had to be applied to high-multiplicity channels.
By contrast, the \FEI addresses the combinatorics problem by performing best-candidate selections during the
reconstruction of the decay chain instead of fixed cuts.
As a consequence, for each event and each decay channel, the \FEI processes the same number of candidates
in vertex fitting and classifier inference i.e. consumes similar amounts of CPU time.
Moreover, the maximum memory consumption is limited due to the fixed number of best-candidates per event, which is a key requirement for using the computing infrastructure.

Finally, the \FEI uses \texttt{FastBDT} \citep{FastBDT}, a gradient-boosted decision tree (BDT) implementation, as its default multivariate classification algorithm.
The algorithm was originally designed for the \FEI to speed up the training and application-phase.
Compared to other popular BDT implementations such as those provided by \texttt{TMVA}~\cite{Hocker:2007ht}, \texttt{SKLearn}~\cite{scikit-learn} and \texttt{XGBoost}~\cite{DBLP:journals/corr/ChenG16}
it originally improved the execution time by more than one order of magnitude, both in training and application. In addition, an improved classification quality was observed. Most of the time when using \texttt{FastBDT} is spent during the extraction of the necessary features, therefore no further significant speedups can be achieved by employing a different method.

\subsection{Automatic Reporting}

The \FEI includes an automatic reporting system called \texttt{Full Event Interpretation Report (FEIR)}. 

The \texttt{FEIR} contains efficiencies and purities for all particles and decay channels at different points during the reconstruction.
Individual reports containing control-plots for each multivariate classifier and input variables are also automatically created.
Control-plots include \textbf{r}eceiver \textbf{o}perating \textbf{c}haracteristics (ROC) curves, which show the tag-side efficiency against purity.
Additionally, for each classifier the purity is plotted as a function of classifier output, to check for a linear relationship as this confirms the classifier output can be treated as a probability.
This built-in monitoring capability upgrades the \FEI from a black-box to a white-box algorithm, which the user can understand and inspect on all levels of reconstruction.

\section{Previous work}
\label{sec:previous_work}
Previous experiments have already developed and successfully employed tagging algorithms.
In order to compare the algorithms, the maximal achievable tag-side efficiency is of particular interest,
because it is directly related to the signal selection efficiency of the measurement.
On the other hand the achievable tag-side purity is only of limited use, because the achievable final purity of the final selection used for the measurement is dominated by the completeness constraint. Hence, most of the incorrect tags can be easily discarded and the final purity depends strongly on the considered signal decay channel.
Moreover, signal-side independent ROC curves are not available for most of the previously employed algorithms. The area under the ROC curve allows one to compare the performance of the tagging algorithms. 

The BaBar experiment \citep{TheBABAR} used the \texttt{Semi-Exclusive $\PB$ reconstruction} (\texttt{SER}) algorithm for hadronic tagging \citep[Section 7.4.1.1]{PhysicsOfTheBFactories}.
The algorithm used exclusive $\PD$ and $\PDst$ mesons candidates as a seed, and combined those with up to 5 charmless hadrons to form a $\BTag$ without assuming an exclusive $\PB$ decay mode.
The tag-side efficiency and tag-side purity of each $\PB$ decay chain was extracted by fitting the beam-constrained mass \citep[Section 7.1.1.2]{PhysicsOfTheBFactories} spectrum of the constructed $\BTag$ meson candidates. The beam-constrained mass is defined as $M_{\mathrm{bc}} = \sqrt{E_{\mathrm{beam}}^2/c^4 - p_{\PB}^2/c^2}$ where $p_{\PB}$ denotes the three-momentum of the reconstructed \PB meson candidate and $E_{\mathrm{beam}}$ denotes half of the centre-of-mass energy of the colliding electron-positron pair. The maximum hadronic tag-side efficiency achieved by this algorithm was $0.2 \%$ for $\PBzero \APBzero$ and $0.4 \%$ for $\PBplus \PBminus$,
with a tag-side purity around $30 \%$. The tag-side purity could be further increased by rejecting $\PB$ meson candidates from low-purity decay chains.
The semileptonic tag was usually constructed by combining an exclusive $\PD$ or $\PDst$ meson with a lepton. 
The maximum semileptonic tag-side efficiency was typically $0.3 \%$ for $\PBzero \APBzero$ and $0.6 \%$ for $\PBplus \PBminus$ with an unknown tag-side purity.

The Belle experiment \citep{Abashian2002117} used the so-called \texttt{Full Reconstruction} (\FR) algorithm \citep{FullReconstruction} for hadronic tagging \citep[Section 7.4.1.2]{PhysicsOfTheBFactories}.
The \FR introduced an hierarchical approach, which is still used by its successor and is presented in this article (see \autoref{sec:method}).
The tag-side efficiency and tag-side purity was extracted by fitting the beam-constrained mass spectrum of the constructed $\BTag$ meson candidates.
The maximum hadronic tag-side efficiency achieved by this algorithm was $0.18 \%$ for $\PBzero \APBzero$ and $0.28 \%$ for $\PBplus \PBminus$,
with a tag-side purity around $10 \%$. Multivariate classifiers \citep{NeuroBayes} were used to estimate the signal probability of each candidate. 
The tag-side purity could be further increased by requiring a minimal signal probability.
Variants of the \FR were used for semileptonic tagging (see \citep{KirchgessnerMasterarbeit} and \citep{KronenbitterPaper}).
The maximum semileptonic tag-side efficiency was $0.31 \%$ for $\PBzero \APBzero$ and $0.34 \%$ for $\PBplus \PBminus$, with a typical tag-side purity of $5 \%$.

Compared to the previously employed algorithms, the \FEI provides a greater tagging and tag-side efficiency, with a equal or better tag-side purity.
The improvements with respect to the \FR can be attributed equally to the additional decay channels and the new candidate selection criteria.
The reported maximum tag-side efficiencies for the previously used exclusive tagging algorithms are summarized in \autoref{sec:results}, \autoref{tab:fei:performance}.
The stated efficiencies are not directly comparable due to different selection criteria,
e.g.: a threshold on the beam-constrained mass or the deviation of the nominal energy from the reconstructed energy $\Delta E = E_{\mathrm{beam}} - E_{\PB}$ with $E_{\PB}$ denoting the energy of the \PB candidate, best-candidate selections, or cuts on the event shape used to suppress background from non-$\PUpsilonFourS$ events.

\section{Results}\label{sec:results}
The \FEI algorithm was developed for the Belle~II experiment.
In order to quantify the improvements with respect to the previously used \FR algorithm, the \FEI is applied to data recorded by the Belle experiment.
Simulated events and recorded data from the Belle experiment are converted into the new Belle~II data format~\citep[Chapter 2]{feithesis}.
This conversion tool was used to validate the entire Belle~II analysis software and will be described in a separate publication~\cite{B2BIIPaper}.
The remainder of this article focuses on the results obtained for the hadronic tag on data recorded by the Belle experiment.
The results for the semileptonic tag and for Belle~II are based on simulated events and are only summarized briefly.
A detailed validation of the entire algorithm can be found in \citet[Chapter 4]{feithesis}.

\subsection{Hadronic Tag}
\label{subsec:hadronic_tag}
The performance of the hadronic tag provided by the \FEI using simulated and recorded Belle events is studied and compared to the previously used \FR algorithm.

At first, the considered decay channels of the \FEI are restricted to the set of hadronic decay channels used by the \FR.
The performance of the \FEI to the \FR are compared using the same hardware and the same simulated charged (neutral) $\PB \APB$ Belle events.
The \FEI required $33 \%$ less computing time and achieved a maximum tag-side efficiency of $0.53 \%$ ($0.33 \%$) on simulated events,
which is significantly higher than the previously reported tag-side efficiencies (see \autoref{sec:previous_work}).
The increase in the maximum tag-side efficiency is due to the improved candidate selection criteria, in particular the best-candidate selections.

Secondly, all decay channels of the \FEI are used, including the $38$ additional hadronic decay channels.
The performance of the \FEI to the \FR using the same hardware and the same simulated charged (neutral) Belle events are then compared. The \FEI required $48 \%$ more computing time and achieved a maximum tag-side efficiency of $0.76 \%$ ($0.46 \%$) on simulated events. The further increase in the maximum tag-side efficiency is due to the additional decay channels.

As mentioned before the maximum tag-side efficiency is an important performance indicator for 
exclusive measurements, which can employ the completeness constraint to achieve a high final purity.
The achieved maximum tag-side efficiencies are summarized in \autoref{tab:fei:performance}.

\begin{figure}
\centering
\includegraphics[width=0.5\textwidth]{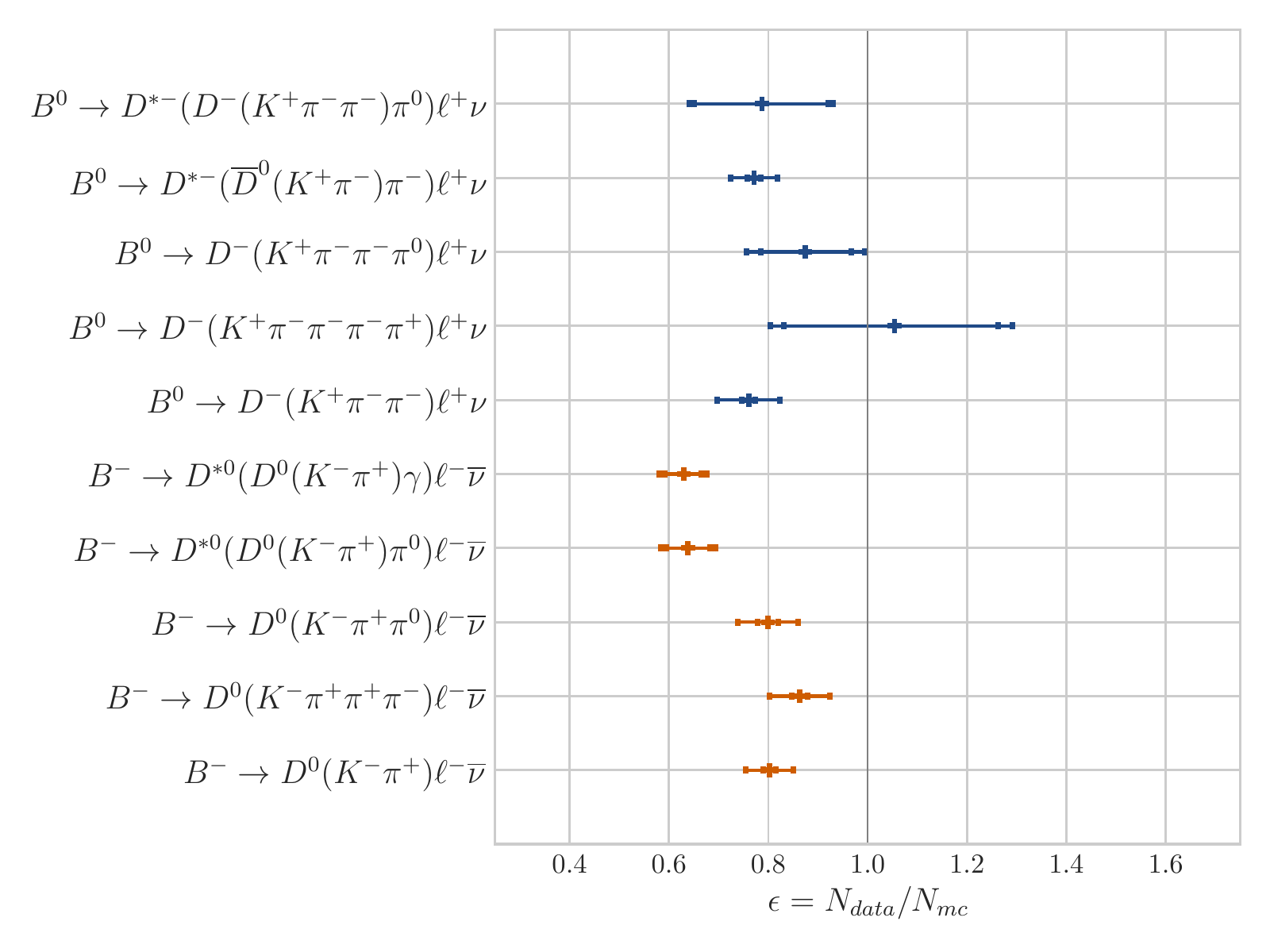}
\caption{The ratios calculated by measuring $10$ semileptonic decay channels on converted Belle data using the hadronic tag. The procedure is described in \citet{Judith}.
}
\label{fig:fei:judith_calibration}
\end{figure}

In order to validate the results for the hadronic tag obtained from the simulation study, we conducted exclusive measurements
of ten different semileptonic  $\PB$ decay channels using the full $\PUpsilonFourS$ dataset recorded by Belle.
The branching fractions of the considered semileptonic decay channels are well-known from independent untagged measurements.
The branching fraction of those well-known decay channels is measured using the hadronic tag, taking into account all known disagreements between simulation and data, e.g. in the particle identification performance and the track reconstruction efficiency. We assume that the remaining disagreement between simulation and data is caused by the tag-side.
Therefore, the ratio $\varepsilon$ of the measured and the expected branching fraction is proportional
to the ratio of the tag-side efficiency on recorded data and simulated events.
Our assumption is supported by the compatibility of the extracted ratios within their uncertainties.
\autoref{fig:fei:judith_calibration} summarizes the results for the ten decay channels.
The ratios averaged over all control-channels for the charged and neutral $\BTag$ mesons are
\begin{align*}
& & \varepsilon_{\mathrm{charged}} &= 0.74^{+0.014}_{-0.013} \pm 0.050 \\
& & \varepsilon_{\mathrm{neutral}} &= 0.86^{+0.045}_{-0.050} \pm 0.054,
\end{align*}
where the first uncertainty is statistical and the second systematic. The systematic uncertainties arises from the signal-side, e.g. through uncertainties on the particle identification performance or the track reconstruction efficiency. 

A detailed description of the control measurements, including results for each tag- and control-channel, can be found in \citet{Judith}.
A similar study was conducted in the past for the \FR by \citet{frcalibration}, yielding a similar overall ratio of $\varepsilon_{\mathrm{comb.}} = 0.75 \pm 0.03$.
The rather large discrepancy between simulated events and recorded data is caused by the uncertainty on the branching fractions
and decay models of the simulated $\PB$ decay channels used for the tag-side and the large number of multivariate classifiers involved in the process.

The uncertainty on the tag-side efficiency of the \FEI is one of the most important
systematic uncertainties in the measurement of branching fractions of rare decays.
The tag-side efficiency can be corrected using the extracted ratios. It is possible
to apply this correction as a function of the tag-side decay channel and signal-probability.
A measurement which uses the ratios to correct the tag-side efficiency is performed relative to the considered calibration decay channels.
The systematic uncertainty of the correction is given by the uncertainty of the ratios.

In order to compare the hadronic tag provided by the \FEI and the \FR in a well-defined manner, which is independent of the signal-side, both algorithms are applied to the same set of $10$ million events. These events are randomly sampled from the full $\PUpsilonFourS$ dataset of $772$ million events recorded by the Belle experiment.
After the tag-side reconstruction, only $\PB$ meson candidates are kept, which fulfill cuts on the beam-constrained mass of $M_{\mathrm{bc}} > \unit[5.24]{GeV}$ and on the deviation of the reconstructed energy from the nominal energy of $\unit[-0.15]{GeV} < \Delta E < \unit[0.1]{GeV}$ calculated on the candidate. In addition, a best-candidate selection is performed, taking the $\PB$ meson candidate with the highest signal probability in each event.

The same cuts on the beam-constrained mass $M_{\mathrm{bc}} > \unit[5.24]{GeV}$ and the deviation of the reconstructed energy from the nominal energy $\unit[-0.15]{GeV} < \Delta E < \unit[0.1]{GeV}$ were applied and only the best (i.e. the highest signal probability) $\PB$ meson candidate in each event was used.

From this dataset, we determined the tag-side efficiency and tag-side purity for different cuts on the signal probability.
We followed the procedure established in previous publications \citep[Chapter 7.1]{PhysicsOfTheBFactories}.
For different cuts on the signal probability, extended unbinned maximum likelihood fits of the beam-constrained mass spectrum are performed. The signal peak consisting of correct $\BTag$ mesons is modelled with a Crystal Ball function \cite{CrystallBall}, whereas the background is described using an ARGUS function \cite{Argus}.
The Gaussian mean of the Crystal Ball function was fixed to the $\PB$ meson mass and its power law exponent was fixed to $m = 4$ based on the expected shape obtained from Monte Carlo simulations.
The location and the width of the ARGUS were fixed using the known kinematic end-point of the spectrum.
All other parameters: the normalization of both functions, the width of the Crystal Ball, and the remaining shape parameters of both functions were adjusted by the fit.
The tag-side efficiency and tag-side purity are determined in a window of $\unit[5.27]{GeV} < M_{\mathrm{bc}} < \unit[5.29]{GeV}$ using the fitted yields of the signal and background component.

In addition, we checked for a potential peaking combinatorial background component, which would bias the results. This test was done using $10$ million events recorded $\unit[60]{MeV}$ below the $\PUpsilonFourS$ resonance. This dataset does not contain $\PB$ mesons, hence no signal is expected. The fitted signal yields were compatible with zero.

\begin{figure}
	\includegraphics[width=0.5\textwidth]{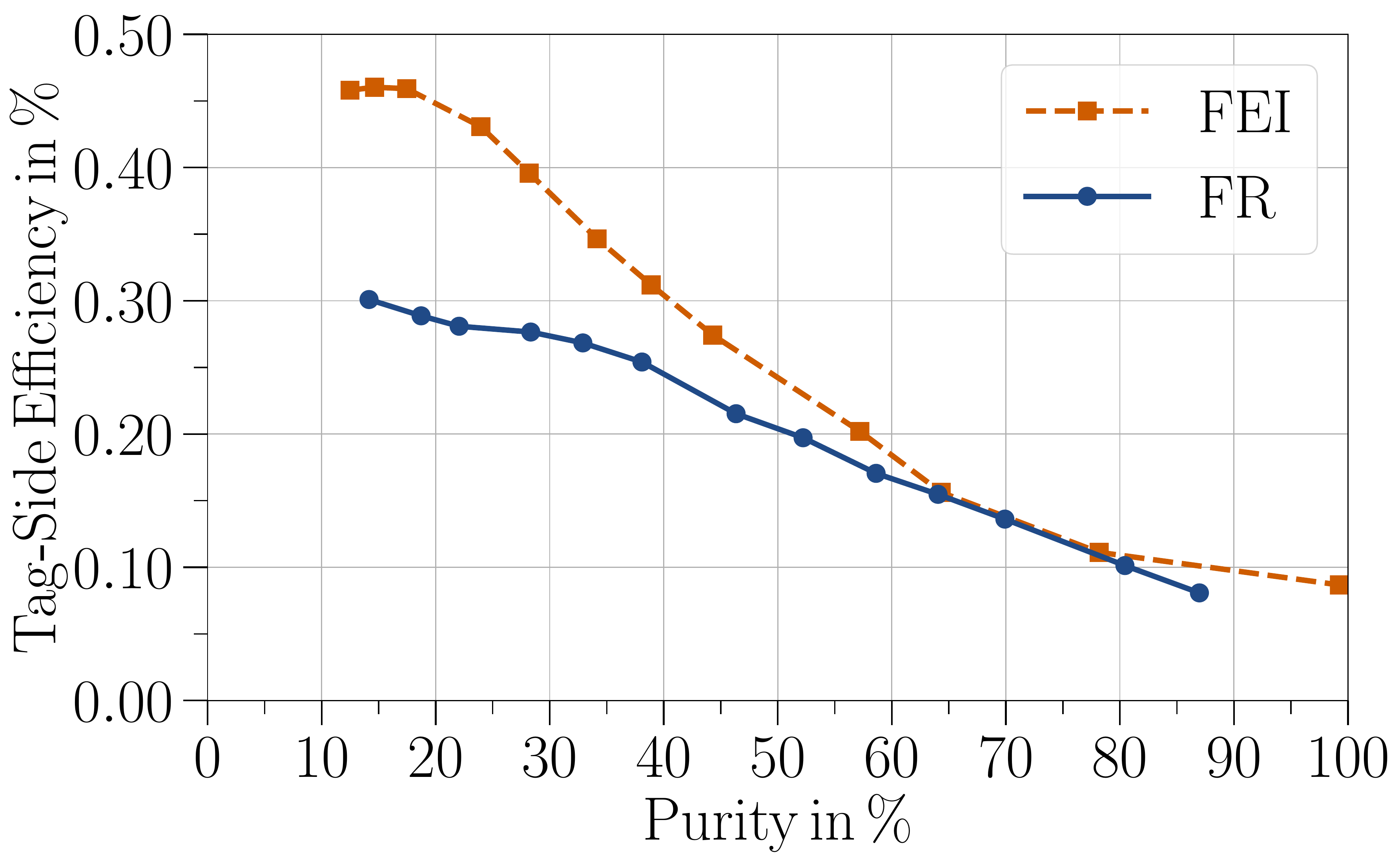}
	\caption{Receiver operating characteristic of charged $\BTag$ mesons extracted from a fit of the beam-constrained mass on converted Belle data. The \FEI outperforms the \FR algorithms performance at low and high purity. }
	\label{fig:roc_charged}
\end{figure}

\begin{figure}
	\includegraphics[width=0.5\textwidth]{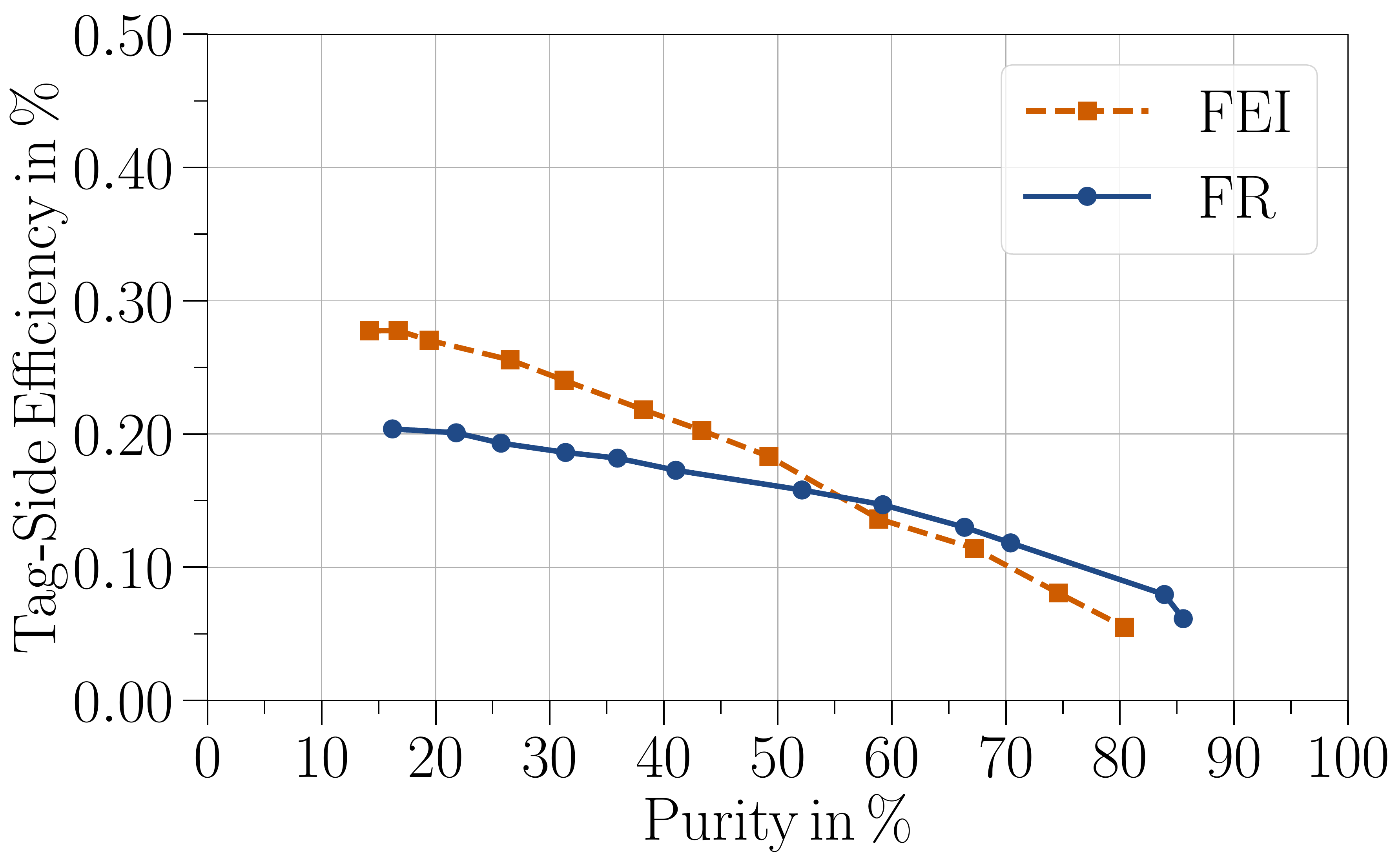}
	\caption{Receiver operating characteristic of neutral $\BTag$ mesons extracted from a fit of the beam-constrained mass on converted Belle data. The \FEI outperforms the \FR algorithms performance at low and intermediate purity. At high purity the tag-side efficiency cannot be extracted reliably.}
	\label{fig:roc_neutral}
\end{figure}

\begin{figure}
	\includegraphics[width=0.5\textwidth]{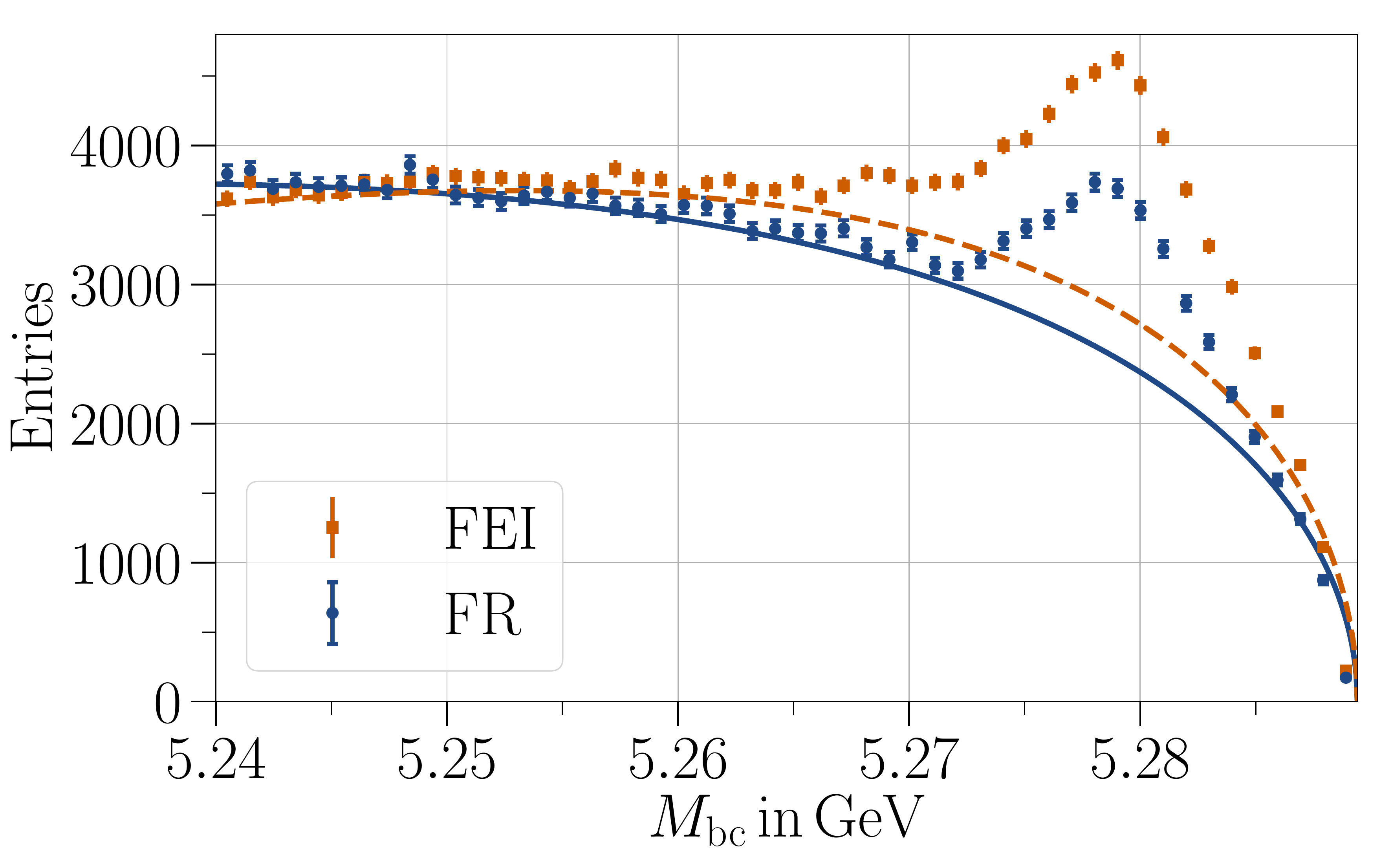}
	\caption{Beam-constrained mass distribution of charged $\BTag$ mesons in the low tag-side purity region on converted Belle data.}
	\label{fig:mbc_charged_loose}
\end{figure}

\begin{figure}
	\includegraphics[width=0.5\textwidth]{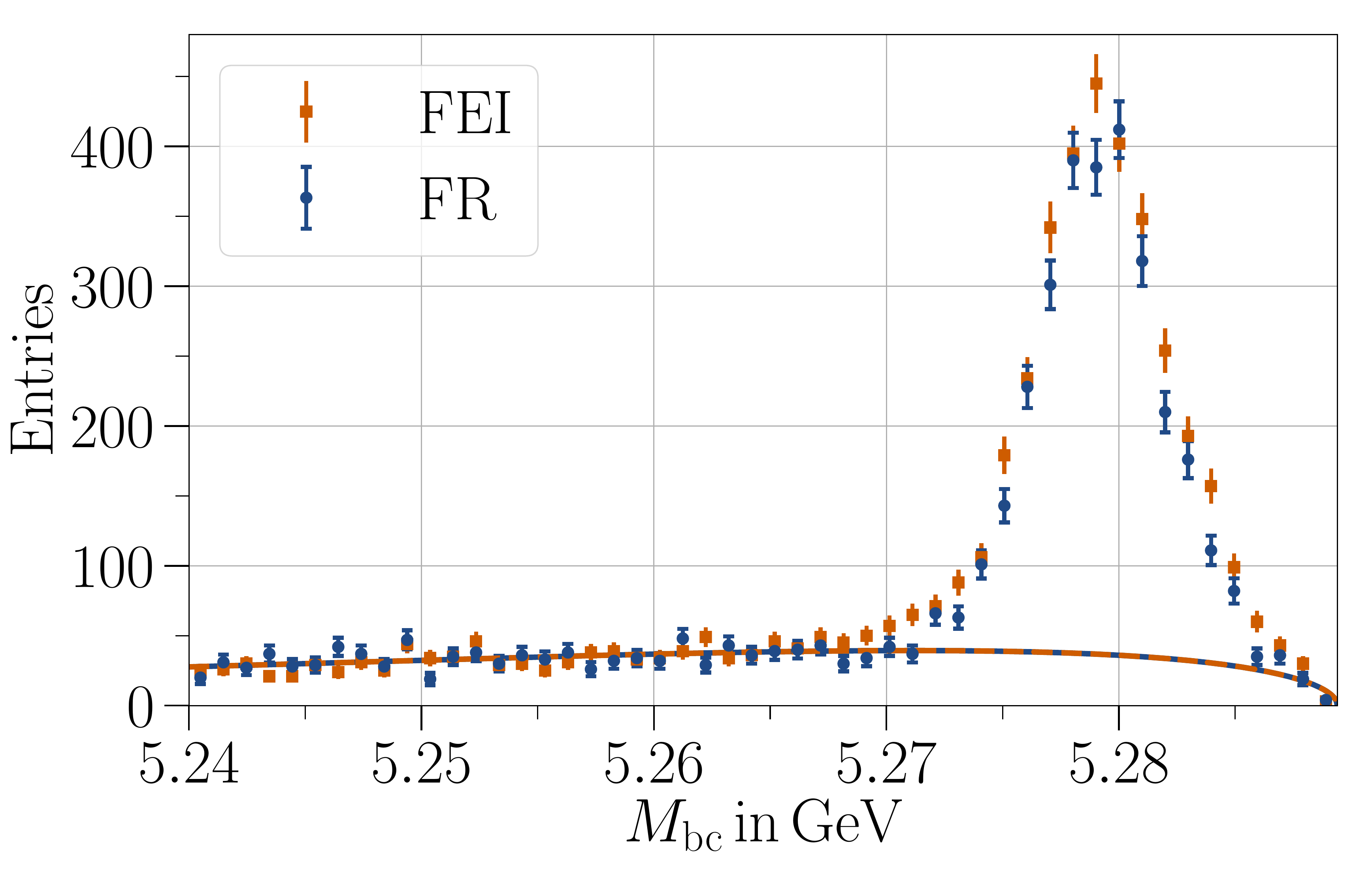}
	\caption{Beam-constrained mass distribution of charged $\BTag$ mesons in the high tag-side purity region on converted Belle data.}
	\label{fig:mbc_charged_hard}
\end{figure}

The resulting ROC curves are shown in \autoref{fig:roc_charged} and \autoref{fig:roc_neutral} for
charged and neutral $\BTag$ mesons respectively.
The \FEI exhibits a larger overall tag-side efficiency compared to the \FR.
We observe a slightly better performance for the \FR than reported in \citet{FullReconstruction}.
Both algorithms perform equally well when requiring a high tag-side purity. We suspect this is because there are only a finite number of cleanly identifiable $\BTag$ meson candidates and both algorithms identify them with similar performance. The results for tag-side purities above $70 \%$ cannot be extracted reliably and depend strongly on the chosen signal or background fit-model. For practical applications, the low tag-side purity regions is of particular interest for exclusive measurements. 
The beam-constrained mass distributions corresponding to the low-purity region with about $15 \%$ tag-side purity and the high-purity region with approximatively $80 \%$ tag-side purity are shown in \autoref{fig:mbc_charged_loose} and \autoref{fig:mbc_charged_hard}, respectively, for the charged $\BTag$.

The maximum tag-side efficiency on recorded data is not determinable by this method, as the fits are restricted to the best $\BTag$ candidates. However, a significant contribution to the improvement of the \FEI compared to the \FR is the increased number of provided candidates per event.
A physics measurement will benefit from these additional tag-side candidates by first combining them with potential signal-side-candidates, applying the completeness constraint (i.e. requiring no additional tracks in the event), and performing the best $\BTag$ candidate selection as the final step of the selection procedure. 
This procedure was successfully used by several measurements to validate the expected improvements on recorded data: \cite{feithesis,FelixMetzner,Judith}.

\subsection{Semileptonic Tag}
The performance of the semileptonic tag provided by the \FEI is studied using simulated Belle events. 
The maximum tag-side efficiencies are summarized in \autoref{tab:fei:performance}.
Receiver operating characteristics extracted from simulated events can be found in \citet{feithesis}.
The results obtained from simulated events, and the fact that the hadronic and semileptonic tag only
share five out of six reconstruction stages, indicate a significant increase in the maximum tag-side efficiency.
The semileptonic tag was successfully used by \citet{feithesis} to determine the branching fraction
of $\PB \rightarrow \Ptau \Pnut$ on the full $\PUpsilonFourS$ dataset recorded by the Belle experiment, 
with a smaller relative statistical uncertainty than obtained previously.
However, no studies with well-known calibration channels as described in \citet{KronenbitterThesis} and no signal-side independent determination of the ROCs as described in \citet{KirchgessnerMasterarbeit}, are available yet.

\subsection{Outlook for Belle~II}
As the Belle II reconstruction software is still being optimized and no large recorded experimental data set was available at the time of writing, hence the final tag-side efficiency cannot be determined reliably for Belle II at this point. Preliminary results can be found in \citep{feithesis} which indicate a worse overall performance. This is likely due to the increased beam background caused by the higher luminosity of the collider, which does lead to additional tracks and neutral energy depositions. This additional detector activity is not yet fully rejected by the Belle II reconstruction algorithms \citep{feithesis} and future improvements are likely possible.

\begin{table}
	\centering
	\caption{Summary of the maximum tag-side efficiency of the Full Event Interpretation and for the previously used exclusive tagging algorithms. For the \FEI simulated data from the last official Monte Carlo campaign of the Belle experiment were used. The maximum tag-side efficiency on recorded data is lower (see \autoref{subsec:hadronic_tag}). The numbers for the older algorithms (see \autoref{sec:previous_work}), are not directly comparable due to different selection criteria, like best-candidate selections and selections to suppress non-$\PUpsilonFourS$ events. 
	}
	\label{tab:fei:performance}
	\begin{tabular}{lrr}
		\toprule
			\belowrulesepcolor
			\rowcolor{HeadRowColor}
		 & $\PBpm$ & $\PBzero$ \\ 
			 \aboverulesepcolor \midrule
			 \multicolumn{3}{c}{Hadronic} \\
			  \midrule
		    \FEI with \FR channels	 &  $0.53$ \% &  $0.33$ \% \\
			\FEI &  $0.76$ \% &  $0.46$ \% \\ \hline
			\FR & $0.28$ \% & $0.18$ \% \\
			\texttt{SER}  & $0.4$ \% & $0.2$ \% \\			
			\midrule
		 \multicolumn{3}{c}{Semileptonic} \\
	  \midrule
		\FEI &  $1.80$ \% &  $2.04$ \% \\  \hline
		\FR &  $0.31$ \% & $0.34$ \% \\
		\texttt{SER} &  $0.3$ \% & $0.6$ \% \\ \bottomrule
	\end{tabular}
\end{table}

%

\section{Discussion}

The multivariate classifiers used by the \FEI are trained on Monte Carlo simulated events.
Depending on the training procedure and the type of events provided to the training, the
multivariate classifiers of the \FEI are optimized for different objectives.

In this article, we presented a so-called \textbf{generic} adaption of the \FEI.
The generic refers to that the \FEI was trained independently of any specific signal-side
using 180 million simulated $\PUpsilonFourS$ events.
This setup optimizes the tag-side efficiency of a ``generic'' $\PUpsilonFourS$.

Other versions of the \FEI exist which optimize the tag-side efficiency
of specific signal events like $\PB \rightarrow \Ptau \Pnu$.
The so-called \textbf{specific} \FEI is trained on the remaining tracks and clusters
after a potential signal $\PB$ meson was already identified.
The training uses simulated $\PUpsilonFourS$ events and simulated signal events.
As a consequence, the classifiers can be specifically trained to identify correctly reconstructed $\BTag$ mesons
for signal events and can focus on reducing non-trivial background which is not discarded by the completeness constraint.
The specific \FEI was first introduced as a proof of concept by \citet{TKeckMasterarbeit} and used in \citet{FelixMetzner}.

Roughly half of the improvements with respect to the previous algorithm can be attributed to the additionally considered decay channels.
Future extensions are currently investigated which use semileptonic $\PD$ meson decays, baryonic decays and decays including $\PKlong$ particles.

It should also be noted that the \FEI algorithm can be applied, with little modification, to the $\PUpsilonFiveS$ resonance. This resonance decays into a pair of $\PB^{(*)} \PB^{(*)}$ and $\PsB^{(*)} \PsB^{(*)}$ mesons.
The powerful completeness constraint can still be applied in this situation.

\section{Conclusion}

The \texttt{Full Event Interpretation} is a new exclusive tagging algorithm developed for the Belle~II experiment that will be used to measure a wide range of decays with a minimum of detectable information. The algorithm exploits the unique setup of $\PB$ factories and significantly improves the tag-side efficiency compared to its predecessor algorithms.

The tag-side efficiency for hadronically tagged $\PB$ mesons was validated and calibrated using Belle data.
Furthermore, the hadronic and the semileptonic tag provided by \FEI have already been used in several validation measurements~\cite{feithesis,BelleFEIlnuGammaPaper,Judith} using the full $\PUpsilonFourS$ dataset recorded by the Belle experiment. Similar studies and measurements for Belle~II are anticipated as soon as the experiment records a sufficient amount of collision events.

There are several ways that the \FEI algorithm could be further refined and applied to so far unexplored applications. These will provide an exciting and fruitful area of future research. 

\section{Acknowledgments}
We thank the KEKB accelerator group, the Belle collaboration, and the Belle~II collaboration for the provided
data and infrastructure.
This research was supported by: the Federal Ministry of Education and Research of Germany (BMBF), the German Research Foundation (DFG),
the Doctoral School ``Karlsruhe School of Elementary and Astroparticle Physics: 
Science and Technology'' funded by the German Research Foundation (DFG), and
the DFG-funded Research Training Group ``GRK 1694: Elementary Particle Physics at Highest Energy and highest Precision''.
F.B. and W.S. are supported by DFG Emmy-Noether Grant No. BE 6075/1-1

\bibliographystyle{unsrtnat}
\bibliography{short}

\end{document}